\documentclass[a4paper]{article}
\usepackage{Odyssey2024}
\usepackage{epsfig,amssymb,amsmath}
\usepackage{booktabs}
\usepackage{color}
\usepackage{xurl}
\usepackage{comment}
\usepackage{nccmath}
\usepackage {colortbl,array,xcolor}
\usepackage{subcaption}
\usepackage{balance}
\usepackage{multirow}
\usepackage{multicol, makecell}
\usepackage{appendix}
\newcommand*\at{\mathbf{a}}
\newcommand*\ee{\mathbf{e}}
\newcommand*\xx{\mathbf{x}}

\newcommand*\zz{\mathbf{z}}
\newcommand*\eepsilon{\boldsymbol{\epsilon}}
\newcommand*\mmu{\boldsymbol{\mu}}
\newcommand*\ssigma{\boldsymbol{\sigma}}
\newcommand{\diag}{\operatorname{diag}}

\usepackage[normalem]{ulem}

\usepackage{adjustbox}
\usepackage{blindtext}

\ninept

\definecolor{Gray}{gray}{0.9}

\title{Do End-to-End Neural Diarization Attractors Need to  \\ Encode Speaker Characteristic Information?}

\makeatletter
\def\name#1{\gdef\@name{#1\\}}
\makeatother
\name{{\em Lin Zhang$^1$\thanks{The work was done during Lin Zhang's visit at BUT-Speech@FIT.}, Themos Stafylakis$^{2,3}$, Federico Landini$^4$,}
     {\em Mireia Diez$^4$, Anna Silnova$^4$, Lukáš Burget$^4$}}

\address{  $^1$National Institute of Informatics, Tokyo, Japan 
  $^2$Omilia - Conversational Intelligence, Athens, Greece \\
  $^3$Athens University of Economics and Business, Athens, Greece \\
  $^4$Brno University of Technology, Faculty of Information Technology, Speech@FIT, Czechia }

\begin{document}

\maketitle

\begin{abstract}
In this paper, we apply the variational information bottleneck approach to end-to-end neural diarization with encoder-decoder attractors (EEND-EDA). This allows us to investigate what information is essential for the model. EEND-EDA utilizes attractors, vector representations of speakers in a conversation. Our analysis shows that, attractors do not necessarily have to contain speaker characteristic information. 
On the other hand, giving the attractors more freedom to allow them to encode some extra (possibly speaker-specific) information leads to small but consistent diarization performance improvements. Despite architectural differences in EEND systems, the notion of attractors and frame embeddings is common to most of them and not specific to EEND-EDA. We believe that the main conclusions of this work can apply to other variants of EEND. Thus, we hope this paper will be a valuable contribution to guide the community to make more informed decisions when designing new systems.
\end{abstract}

\section{Introduction}

End-to-end speaker diarization has gained considerable popularity in the last few years with several works exploring different frameworks~\cite{park2022review}. Among them, end-to-end neural diarization (EEND)~\cite{fujita_end--end_2019} is one of the most studied ones. It formulates diarization as a per-speaker per-frame binary classification problem where the predictions are represented by a two-dimensional matrix across time and speaker. Unlike modular diarization systems, this formulation effectively covers all aspects of the task with a single system including voice activity detection and handling of overlapped speech.

Several extensions have been proposed to this framework, among which EEND with encoder-decoder-based attractors (EEND-EDA)~\cite{horiguchi_encoder-decoder_2022} was the first successful variant capable of dealing with a variable number of speakers. In EEND-EDA, frame embeddings are produced for the whole recording and these are encoded into a single representation from which attractors are decoded. Each attractor represents one speaker and they are compared by means of dot-product with the frame embeddings to determine which speakers are active at which frames.

EEND-EDA has become the main variant capable of handling multiple speakers and many modifications or alternatives have been proposed. In the original version, long short-term memory (LSTM) layers are used for the encoder and decoder. The decoder's initial hidden state is initialized with the final hidden state of the encoder and the decoder is fed with 0's to decode attractors one by one. Different ideas have been proposed to improve the attractor-decoding scheme. Some works~\cite{pan_towards_2022,broughton_improving_2023} have improved the model performance by feeding the LSTM decoder with global representations drawn from the frames rather than uninformative 0's. Other works~\cite{rybicka22_interspeech,fujita2023neural,hao2023nn,chen_attention-based_2023,chen2024TASLP, landini_diaper_2023,harkonen2024eend} have proposed replacing the encoder-decoder module by an attention-based scheme using different approaches. All these methods still rely on the notion of ``attractors'' which are compared with frame embeddings through dot-product. Nevertheless, little analysis has been presented on the attractors and no study until now has focused on understanding what information is needed to be encoded in them. While in principle attractors are assumed to encode speaker information necessary to identify which frames were ``spoken'' by a certain speaker, there is no evidence of this or up to which extent they need to encode speaker characteristic information.

In information theory, the concept of an information bottleneck (IB) \cite{Tishby1999ib} was introduced to enforce the encoding to act as a minimal sufficient statistic of the input for predicting the target. To optimize the IB objective for deep neural networks, the deep variational information bottleneck (VIB) \cite{alemi2017vib} was proposed. VIB utilizes variational inference to construct a lower bound of the IB objective, which enables optimization through stochastic gradient descent. 

Given its capability to capture essential information within neural networks, we applied VIB into  EEND-EDA\footnote{\url{https://github.com/BUTSpeechFIT/EENDEDA_VIB}} to better understand the end-to-end diarization mechanism. Specifically, we replace the point estimates of the frame embeddings and attractors of the original EEND-EDA with normal distributions with parameters estimated by VIB.
Our results indicate that it is possible to obtain similar performance to the original EEND-EDA even if the attractors are regularized towards a standard normal distribution, making them less discriminative. This suggests that attractors might not need to encode specific speaker identities but rather enough information to distinguish them in a given conversation. At the same time, slight but consistent improvements can be obtained if the attractors have more freedom, suggesting that some specific speaker information is relevant to the model.

These findings could lead to more effective strategies for training speaker diarization systems and more parameter-efficient models. The analysis shows that competitive end-to-end diarization models can be trained even if the frame embeddings and attractors do not contain much speaker-specific information. While it was not the main focus of this work, the strategies we utilized can also be used if privacy concerns are relevant.

\begin{figure*}[!h]
    \centering
     \begin{subfigure}[b]{0.43\textwidth}
         \centering
         \includegraphics[width=\textwidth]{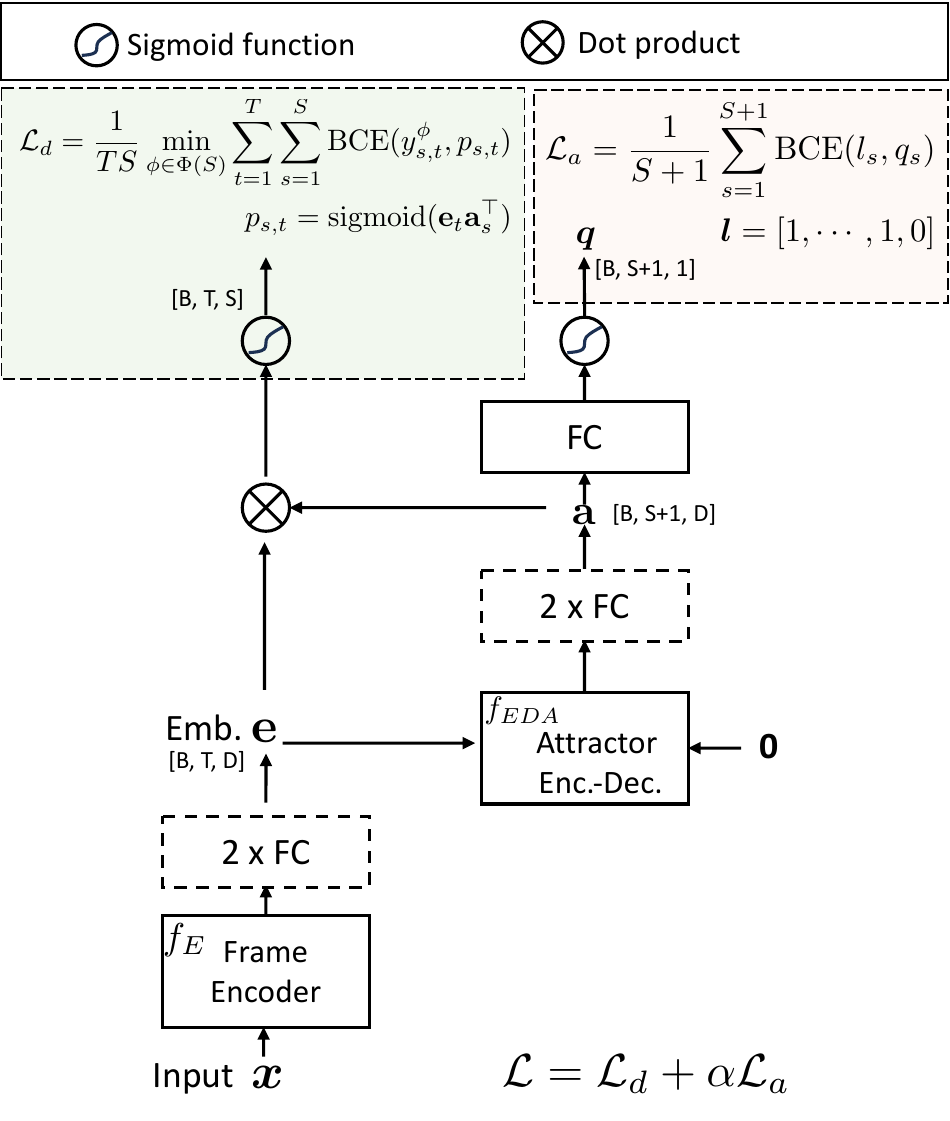}
         \caption{\it Original EEND-EDA}
         \label{fig:ori_eend-eda}
     \end{subfigure}
     \hfill
     \begin{subfigure}[b]{0.53\textwidth}
         \centering
         \includegraphics[width=\textwidth]{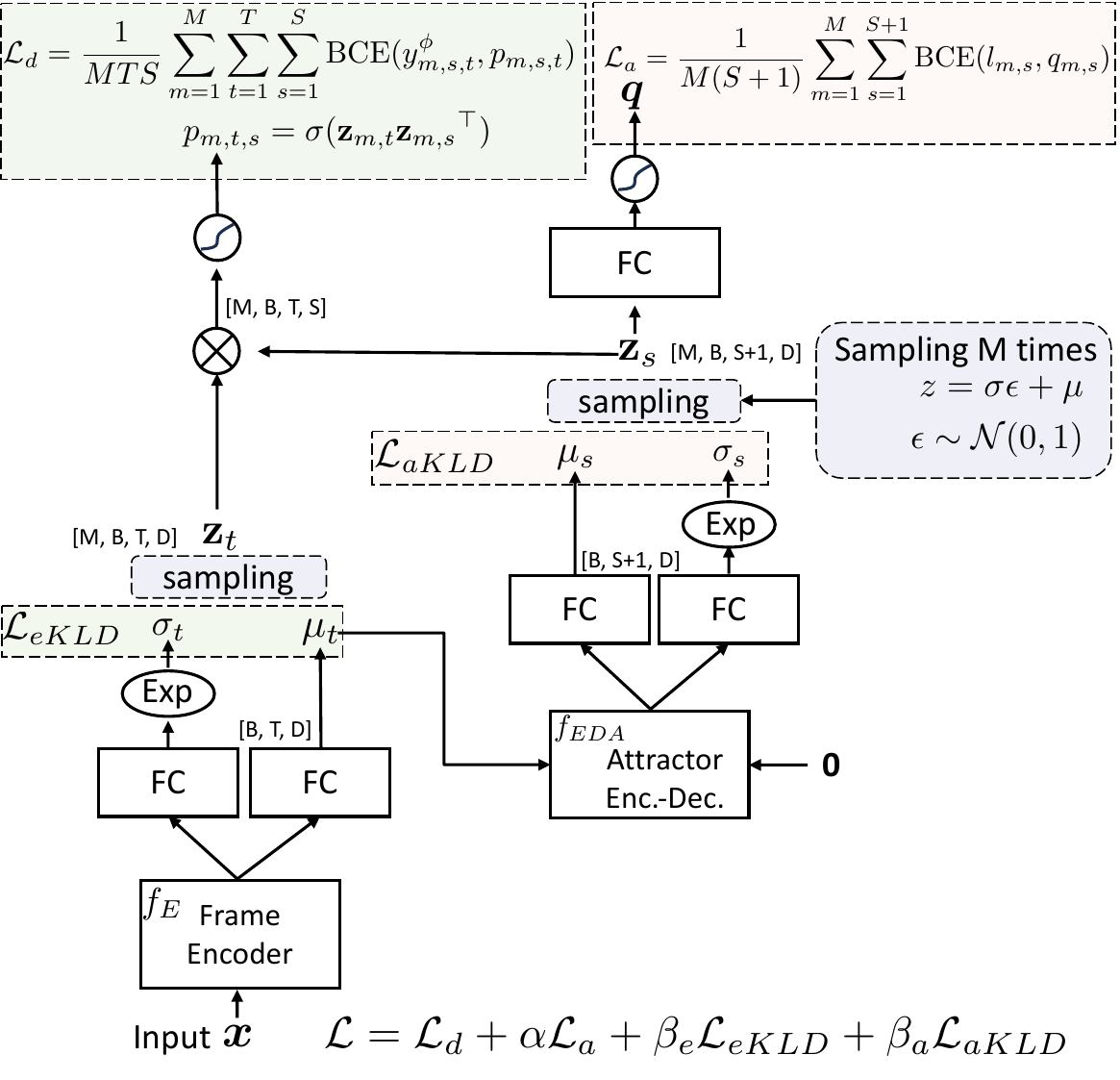}
         \caption{\it EEND-EDA with VIB}
         \label{fig:eeda-eda-vib}
     \end{subfigure}
     \vspace{-2mm}
    \caption{\it Comparison between the original EEND-EDA model and EEND-EDA with VIB. Numbers within [] present dimensions. M is the sampling number, B is batch size, T is number of frames, D is feature dimension, and S is the maximum speaker number within one mini-batch. In (a), four additional FC layers in the dashed box are introduced to make the original EEND-EDA comparable with (b).}
    \label{fig:eend_eda}
    \vspace{-2mm}
\end{figure*}

\section{EEND-EDA with VIB}

\subsection{EEND with Encoder-Decoder-Based Attractors}\label{sec:eendeda}

As shown in Fig. \ref{fig:ori_eend-eda}, EEND-EDA \cite{horiguchi_encoder-decoder_2022} can be viewed as a combination of two branches: a frame embedding branch and an attractor branch.
The frame embedding branch, uses a self-attention encoder $f_\text{E}(\cdot)$ to convert the $D$-dimensional input acoustic feature for $T$ frames $\boldsymbol{x}_{1:T}$ into frame embeddings:
\begin{equation}\label{eq:eendeda_e}
\begin{aligned}
    \boldsymbol{\ee}_{1:T}  &=  f_\text{E}(\boldsymbol{\xx}_{1:T}).\\
\end{aligned}
\end{equation}

The attractor branch uses an encoder-decoder module $f_\text{EDA}(\cdot)$, usually consisting of two unidirectional LSTMs, to decode attractors (each one representing one speaker):
\begin{equation}
    \begin{aligned}
        \at_{1:(S+1)} &= f_{EDA}(\ee_{1:T}).
    \end{aligned}
\end{equation}

Then, to determine the number of speakers present in the input recording, it estimates the existence probability of each attractor $s$ as:
\begin{equation}
    q_s = \text{sigmoid}(\text{FC}(\at_s)), s\in [1:S+1], \\ 
    \label{eq:attractor_ex_prob}
\end{equation}
where $\text{FC}$ denotes a fully connected linear layer.

During training, the attractor loss is defined:
\begin{equation}\label{eq:l_a}
    \mathcal{L}_a = \frac{1}{S+1}\sum_{s=1}^{S+1}\text{BCE}(l_s, q_s),
\end{equation}
\noindent where $S$ is the number of valid attractors (corresponding to the number of speakers), $\boldsymbol{l}_{1:(S+1)} = [1, \cdots, 1, 0]$ is the attractor validity label and $\text{BCE}(\cdot, \cdot)$ is the binary cross entropy function.
During inference, 
attractors are sequentially extracted until $q_s$ is below a predefined threshold $\tau$, usually set as $0.5$. 

Finally, the dot product between the per-frame embeddings and the attractors  is used to produce the diarization results:

\begin{equation}\label{eq:eendeda_p}
\begin{aligned}
    p_{s,t}  &= \text{sigmoid}(\ee_t\at_s^\top),\\
\end{aligned}
\end{equation}

\noindent where $(\cdot)^\top$ denotes the matrix transpose and $\boldsymbol{p} = \{p_{s,t}|s\in[1,S], t\in [1,T]\}$ is a matrix of per-frame per-speaker activity probabilities. Given the ground-truth speech activities $y_{s,t}$, where $y_{s,t} = 1$ when the $s$-th speaker is active in the $t$-th frame and $y_{s,t} = 0$ otherwise, the diarization loss is calculated using permutation invariant training \cite{Yu2017PIT}:
\begin{equation}\label{eq:l_d}
    \mathcal{L}_{d} = \frac{1}{TS}\min_{\phi \in \Phi(S)} \sum_{t=1}^{T}\sum_{s=1}^{S}\text{BCE}(y^{\phi}_{s,t}, p_{s,t}).
\end{equation}
where $\min_{\phi \in \Phi(S)}$ is to find the permuted sequence $\phi$ that has the minimum loss among all possible permutations $\Phi$ of $S$ speakers.
The model is trained using a total loss, which is a combination of both defined losses:
\begin{equation}\label{eq:loss_eend-eda}
    \mathcal{L} = \mathcal{L}_{d} + \alpha\mathcal{L}_{a}. 
\end{equation}

\subsection{Variational Information Bottleneck}\label{sec:vib}
Given a distribution $p(x,y) = p(y|x)p(x)$ of inputs $x$ and targets $y$, the aim of the information bottleneck method proposed in~\cite{Tishby1999ib} is to find a stochastic encoding $p(z|x)$ that is maximally informative about the target $y$, and, at the same time, preserves minimum information about the input $x$. This is achieved by finding the encoding that maximizes the difference between mutual informations $I(y, z) - \beta I(x, z)$, where $\beta$ controls the trade-off between $z$ being expressive about $y$ and compressive about $x$. Although, it is intractable to find the optimal encoding for nontrivial distributions $p(x,y)$, Alemi \MakeLowercase{\textit{et al.}} \cite{alemi2017vib} proposed Deep Variational Information Bottleneck (VIB) -- a variational approximation to this problem, where a neural network is trained to estimate the stochastic encoding $p(z|x)$.

\subsubsection{VIB model and objective}
As a simple VIB example, let us first consider the classification task described in~\cite{alemi2017vib}, where the distribution $p(\xx,y)$ is represented by the empirical distribution of the training input observations $\xx_n$ and the corresponding categorical labels $y_n$. For simplicity, the stochastic encoding will be represented by a Gaussian distribution with diagonal covariance matrix $p(\zz|\xx) = \mathcal{N}(\zz;\mmu(\xx), \diag(\ssigma(\xx)^2))$, where the mean vector $\mmu(\xx)$ and vector of standard deviations $\ssigma(\xx)$ are functions of the input $\xx$ and these functions are parametrized by a neural network called {\em encoder}. To train the {\em encoder} network, 
\cite{alemi2017vib} proposes to optimize the following objective 

{\footnotesize
\begin{eqnarray}\label{eq:vib_loss1}
    & \frac{1}{N}\sum_{n=1}^N \int p(\zz|\xx_n)\log q(y_n|\zz) - \beta p(\zz|\xx_n)\log \frac{p(\zz|\xx_n)}{r(\zz)}d\zz  \\
    \label{eq:vib_loss2}
    & = \frac{1}{N}\sum_{n=1}^N \left( \mathbb{E}_{\eepsilon \sim \mathcal{N}(\mathbf{0},\mathbf{I})}\left[-\log q(y_n| \mmu(\xx_n) +  \eepsilon\odot\ssigma(\xx_n) )\right] \right. \nonumber \\
     &\ \qquad\qquad
     \left. + \beta \mathrm{KL}\left[p(\zz|\xx_n)|| r(\zz)\right] \right)
\end{eqnarray}
}

\noindent as a lower-bound to the original information bottleneck objective $I(y, \zz) - \beta I(\xx, \zz)$. The terms $q(y_n|\zz)$ and $r(\zz)$ are variational approximations of $p(y_n|\zz)$ and $p(\zz)$ whose exact evaluation (given the learned $p(\zz|\xx)$ and the empirical $p(\xx,y)$) is intractable. The approximate distribution $q(y_n|\zz)$ is represented by another neural network called {\em decoder}, which maps vector $\zz$ to a (categorical) distribution of $y$. This network is trained jointly with the {\em encoder} network using the same objective \eqref{eq:vib_loss1}. The approximate marginal distribution $r(\zz)$ can also be trained, but in~\cite{alemi2017vib} it is fixed as a standard normal distribution as the learned $p(\zz|\xx)$ can accommodate to it.

\subsubsection{VIB training}
The first term in \eqref{eq:vib_loss1} is the expected cross-entropy evaluated for the decoder output where the expectation is with respect to the encoding distribution $p(\zz|\xx)$. This term clearly favours $p(\zz|\xx)$ that is expressive about $y$ as required by the IB principle. This term also encourages the distribution $p(\zz|\xx)$ to be more certain (having lower variance) peaking around $\zz$ for which $\log q(y_n|\zz)$ is high. The second term is the Kullback-Leibler (KL) divergence between the encoding $p(\zz|\xx)$ and the (approximate) marginal distribution $r(\zz)\approx p(\zz)$ favouring more uncertain and thus less expressive encoding. 
Clearly, in the extreme case when the 
encoding  $p(\zz|\xx)$ equals $p(\zz)$ for any input $\xx$, it does not contain any information about the input. 
As we will see in later experiments, $\beta$ controls the trade-off between the two terms in the objective where low values of $\beta$ lead to more certain encoding distributions $p(\zz|\xx)$ that are more expressive about the output $y$ and higher values of $\beta$ lead to more uncertain distributions that are less expressive about the input $\xx$.

In order to optimize this objective during the training, the first term in \eqref{eq:vib_loss2}, is rewritten using the standard reparameterization trick \cite{kingma2013auto}, which allows to easily back-propagate through it.
Since we use Gaussian distributions for both $p(\zz|\xx)$ and $r(\zz)$, the KL divergence can be expressed analytically and it is therefore also easy to back-propagate through it.

The \emph{encoder} and \emph{decoder} networks can be trained with a variant of stochastic gradient descent where, for each training example $\xx_n$, we forward propagate through the encoder network to obtain the parameters of the encoding distribution $\mmu(\xx_n)$ and $\ssigma(\xx_n)$. 
Next, we sample from the encoding distribution using the reparameterization trick 
\begin{equation}
\label{eq:reparametrization_trick_VIB}
    \hat{\zz}_n = \mmu(\xx_n) +  \eepsilon\odot\ssigma(\xx_n),
\end{equation} 
\noindent where $\eepsilon$ is a sample from the standard normal distribution, and the operator $\odot$ denotes element-wise product.
We forward propagate this sample through the decoder network to evaluate $q(y_n|\hat{\zz}_n)$. This way, we obtain all the quantities necessary for evaluating the objective \eqref{eq:vib_loss2} approximating the expectation in the first term using only a single sample. This is often sufficient for the training, but multiple samples can be used to better approximate the expectation.
Finally, parameters are updated by back-propagating through this objective and the \emph{encoder} and \emph{decoder} networks.

\subsubsection{VIB regularized classifier}
Once the \emph{encoder} and \emph{decoder} networks are trained, we can consider them in a tandem as a single classification network, where  $\zz$ functions as a probabilistic representation at the output of an internal hidden layer. The VIB training makes the hidden representation $\zz$ forget the details of the input $\xx$  which are irrelevant to the output, which can be seen as a specific regularization. It is shown in \cite{alemi2017vib} that VIB training performs competitively with other regularization techniques and produces systems more robust to adversarial attacks.
The disadvantage of such VIB trained classifier is that the uncertainty of the hidden representation $\zz$ should be taken into account during the inference. Therefore, to correctly evaluate the output, the expected probability of $p(y|\xx)$ should be estimated by averaging the decoder output $q(y|\hat{\zz})$ obtained from many samples of $\hat{\zz}$. However, for practical inference only the mean $\mmu(\xx)$ of the distribution $p(\zz|\xx)$ may be used.

\subsection{EEND-EDA regularized using VIB}\label{sec:veendeda}
In our experiments, we will use the VIB principle to regularize the EEND-EDA model, to limit the information encoded in the frame embeddings and attractors. This way, we want to find what is the minimum necessary information encoded in these representations that allows for good diarization performance.
For this purpose, we extend and train the EEND-EDA using the VIB as shown in Fig. \ref{fig:eeda-eda-vib}. Each frame embedding and each attractor are treated as stochastic encodings. 
Each frame embedding $\boldsymbol{e}_t$ is still evaluated as in \eqref{eq:eendeda_e}, but it is further transformed by two fully connected (FC) layers, to obtain the parameters of the stochastic encodings:

\begin{equation}
\begin{aligned}
    \boldsymbol{\mu}_t = \text{FC}^{\mu_t}(\ee_t),  \\
    \boldsymbol{\sigma}_t = \exp(\text{FC}^{\sigma_t}(\ee_t)),  \\
\end{aligned}
\end{equation}
\begin{equation}
\label{eq:frames_enc_VIB}
    p(\zz_t|\xx_{1:T}) = \mathcal{N}(\zz_t|\mmu_t, \diag(\ssigma_t^2)).
\end{equation}

Similarly, we obtain the stochastic encodings for the attractors as:

\begin{equation}
\begin{aligned}
    \boldsymbol{\mu}_s = \text{FC}^{\mu_s}(\at_s),  \\
    \boldsymbol{\sigma}_s = \exp(\text{FC}^{\sigma_s}(\at_s)),  \\
\end{aligned}
\end{equation}

\begin{equation}
\label{eq:attractor_enc_VIB}
    p(\zz_s|\xx_{1:T}) = \mathcal{N}(\zz_s|\mmu_s, \diag(\ssigma_s^{2})).
\end{equation}

To train the EEND-EDA with VIB, we proceed in a similar manner as described in Sec. \ref{sec:vib}: we forward propagate through the encoder network, which is represented by \eqref{eq:attractor_enc_VIB} and \eqref{eq:frames_enc_VIB}. We sample the encodings $\hat{\zz}_s$ and $\hat{\zz}_t$ (using the reparameterization trick, see equation \eqref{eq:reparametrization_trick_VIB}) for each frame and for each attractor. We pass these samples through the decoder represented by \eqref{eq:attractor_ex_prob} and \eqref{eq:eendeda_p}. In these formulas, we only substitute the deterministic attractor embeddings $\boldsymbol{a}_s$ and frame embeddings $\boldsymbol{e}_t$ with the corresponding sampled encodings $\hat{\zz}_s$ and $\hat{\zz}_t$, respectively.

The final loss to optimize is: 
\begin{equation}\label{eq:loss_veend-eda}
\begin{aligned}
    \mathcal{L} &= \mathcal{L}_{d}  + \beta_e\mathcal{L}_{eKLD} + \alpha\mathcal{L}_{a} +  \beta_a\mathcal{L}_{aKLD},
\end{aligned}
\end{equation}
\noindent where $\mathcal{L}_{d}$ and  $\mathcal{L}_{a}$ are defined in \eqref{eq:l_d} and \eqref{eq:l_a}, respectively. $\beta_e$ and $\beta_a$ control the amount of regularization imposed on the encodings by the KLD losses:
\begin{eqnarray}
\label{eq:KLD_loss_e}
        \mathcal{L}_{eKLD} &= \frac{1}{T} \sum_{t=1}^T\text{KL}(\mathcal{N}(\boldsymbol{\mu}_t,\boldsymbol{\sigma}_t^2)||\mathcal{N}(\boldsymbol{0},\boldsymbol{I})), \\
\label{eq:KLD_loss_a}
        \mathcal{L}_{aKLD} &= \frac{1}{S} \sum_{s=1}^{S+1}\text{KL}(\mathcal{N}(\boldsymbol{\mu}_s,\boldsymbol{\sigma}_s^2)||\mathcal{N}(\boldsymbol{0},\boldsymbol{I})).
\end{eqnarray}

In the text above, we consider that the losses $\mathcal{L}_{a}$ and $\mathcal{L}_{d}$ are evaluated using a single sample of each $\zz_s$ and $\zz_t$, which corresponds to approximating the expectation in \eqref{eq:vib_loss2} using only a single sample. 
This is the configuration used in most of our experiments, as it produced good results. 
However, the expectation could also be approximated by multiple samples. In such case, $M$ samples would be generated for each $\zz_s$ and $\zz_t$, and the objectives $\mathcal{L}_{a}$ and $\mathcal{L}_{d}$ would be evaluated $M$ times and averaged.
We present results for these different training strategies in Sec. \ref{sec:selec_best_perm}.
 
Similarly, multiple samples should be also used for the inference with the trained model. In such case, the speaker activity probabilities $p_{s,t}$ and the attractor existence probabilities $q_s$ (see equations \eqref{eq:eendeda_p} and \eqref{eq:attractor_ex_prob}) are evaluated and averaged for multiple samples. Alternatively, as a computationally cheaper approximation, \eqref{eq:eendeda_p} and \eqref{eq:attractor_ex_prob} can be evaluated using the encoding means $\mmu_t$ and $\mmu_s$. This later strategy is used to report all the DER values in this paper, as it is more computationally efficient and we found it to produce similar results.

In our experiments, we also consider configurations where either the frame embeddings or the attractors in the EEND-EDA are stochastic while maintaining the other component as deterministic. This is achieved by setting the weight of the KLD loss to zero and directly using the corresponding means $\mmu_t$ or $\mmu_s$ as the deterministic embeddings. In addition, setting both $\beta_e$ and $\beta_a$ to zero and using only the mean vectors $\boldsymbol{\mu}_t$ and $\boldsymbol{\mu}_s$ recovers the original EEND-EDA.

\section{Experimental Setup}
\subsection{Data}

As usual with the end-to-end neural diarization framework, the models are mostly trained on synthetic data. Following \cite{landini22SC,landini2023multi}, we utilize simulated conversations (SC) to allow better performance than simulated mixtures \cite{fujita_end--end_2019}. Using statistical information derived from the development set of the conversational telephone speech domain from DIHARD-III \cite{ryant2020dihard3}, three different sets of 2480\,h each were generated where recordings had 2, 3, and 4 speakers respectively. The first set, denoted ``SC2'' was used to train the model from scratch. The three sets were pooled to form ``SC2-4'', containing 7442\,h of audio, which was used for a second step of training as described in the next section.

For the evaluation of the models, we use the Callhome \cite{SRE2000callhome} dataset consisting of telephone conversations. This dataset is commonly used to evaluate diarization systems and it is divided into two parts\footnote{\scriptsize\url{https://github.com/BUTSpeechFIT/CALLHOME_sublists}} denoted CH1 (with 8.7\,h and 2-7 speakers) and CH2 (with 8.55\,h and 2-6 speakers). Besides, for some of the comparisons we use the subset of CH1 with only recordings containing two speakers denoted CH1-2spk (with 3.2\,h).

\subsection{Configurations}

The configurations of EEND-EDA are the same as in \cite{horiguchi_encoder-decoder_2022, landini2023multi} and $\alpha$ is set as $1$. Our analyses are based on a previously released PyTorch implementation\footnote{\scriptsize\url{https://github.com/BUTSpeechFIT/EEND}}. The only difference with that EEND-EDA implementation is that, as mentioned in Section \ref{sec:veendeda}, four additional FC layers (each with 256 units) are introduced into the structure. Therefore, to ensure the proposed model is comparable to EEND-EDA, both versions of EEND-EDA, with and without the additional layers, are used as baselines. Specifically, when EEND-EDA has four additional FC layers, two FC layers are inserted after $f_{E}$ and the other two are positioned after $f_{EDA}$ as shown in dashed boxes of Fig.~\ref{fig:ori_eend-eda}. 

Following standard practice when training EEND-EDA, there are three steps. (1) Initial training using SC with two speakers (using SC2) is run for 100 epochs. Then, we average the parameters of the last ten checkpoints every ten epochs, and evaluate performance on CH1-2spk to determine the optimal model. The ten checkpoints are averaged for the subsequent adaptation step. (2) Adaptation to a set with a variable number of speakers (using SC2-4) is run for 80 epochs. Then, we average the parameters of the last ten checkpoints every ten epochs, and evaluate performance on CH1 to determine the optimal model. The ten checkpoints are averaged for the subsequent fine-tuning step. (3) Fine-tuning to in-domain data (using CH1) for 100 epochs. We average the parameters of the last ten checkpoints every ten epochs, and evaluate performance on CH1 to determine the optimal model. Then, such a model is used to report results on CH2.

We use the diarization error rate (DER) as defined by NIST~\cite{NISTRT} for evaluating the performance, considering overlaps and 0.25 collar as is standard practice for Callhome. {Confidence intervals (CI) were calculated using the Interspeech official toolkit \footnote{\scriptsize\url{https://github.com/luferrer/ConfidenceIntervals}} with the default configuration.}

\section{Results and Discussions}

\subsection{Impact of the weight of the KLD loss}\label{sec:res_weight}

\begin{figure}[!th]
    \centering
    \includegraphics[width=\columnwidth]{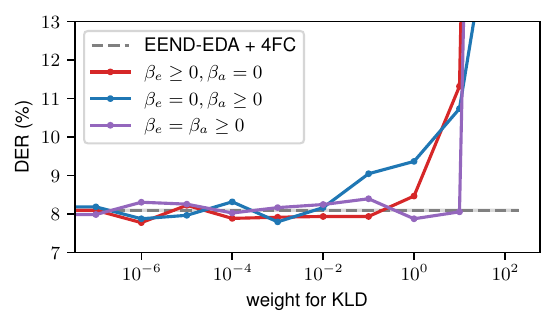}
    \vspace*{-3mm}
    \caption{\it DER (\%) in CH1-2spk for different values of $\beta_e$ and $\beta_a$. EEND-EDA + 4FC with five runs are shown as baselines within the gray shade, and the dashed line represents the mean of those five runs (overlapped due to the small variance).}\label{fig:weight_KLD}
    \vspace*{-5mm}
\end{figure}

In this section, 
we will analyze the effect that different degrees for regularization have in the diarization performance. 
We individually analyze the impact of using $\beta_e$ and $\beta_a$ to control the degree of regularization on frame embeddings and attractors, respectively. For analyzing the effect of regularizing only frame embeddings or only attractors, we use the configurations outlined at the end of Section \ref{sec:veendeda}.
That is, to analyze only the effect of $\beta_e$, we set $\beta_a=0$, and directly use the mean $\mmu_s$ estimated by the network, and vice versa.
We also explore the combination of $\beta_a$ and $\beta_e$. Fig.~\ref{fig:weight_KLD} shows how diarization performance varies with the weights used for the KLD losses. 
We observe that the DER of the model with VIB is virtually the same as the baseline \footnote{
As sharp readers will spot, it is \textit{superfluous} to have two consecutive FC layers, still, we kept them to have consistency in the number of parameters.}
for a wide range of weights. Only relatively large values of $\beta$ result in performance degradation.
We should not expect that the VIB would (significantly) improve the performance as the VIB regularization only limits the information contained in the frame embeddings and attractors.
In section \ref{sec:visualization} we will analyze how the DER correlates with the amount of information that is preserved in these representations.

\subsection{Number of samples and selection of the permutation $\phi$}
\label{sec:selec_best_perm}
As commented in Sec. \ref{sec:veendeda}, for each training example the expectation in \eqref{eq:vib_loss2} can be approximated using only a single or multiple samples of each encoding $z_t$ and $z_s$. 
The first three columns in Table \ref{tab:train_mu_sample} show the performance for 1, 12 and 100 samples, on CH1-2spk using $\beta_e=\beta_a=10^0$ (as it was one of the best-performing configurations seen in Section \ref{sec:res_weight}).
It can be seen that more samples help, however, when training with multiple samples, it is not clear how to choose the permutation $\phi$  for  $\mathcal{L}_{d}$ in \eqref{eq:l_d} since different samples can favor different permutations. In our case, we select the best permutation for each sample, use it to evaluate the corresponding loss, and average the loss of all samples.
The problem with this solution is that it is computationally expensive, as the best permutation has to be found for each sample. At the same time, it is not clear if using potentially a different permutation for each sample can deteriorate the training process. 
Therefore, we also consider the variant where the best permutation is estimated using the speaker activities $p_{s,t}$ obtained from equation \eqref{eq:eendeda_p} evaluated with the
mean encoding vectors $\mmu_t$ and $\mmu_s$. Such permutation should be compatible with any encoding samples.
{The last three columns in Table \ref{tab:train_mu_sample} show the result where this strategy is used to find permutation $\phi$ and different samples are used for training.
Given that their results are similar to each other, we choose the relatively \emph{cheaper} strategy—using $\mu$ to search for the best permutation and one sample to calculate losses—for the subsequent analysis and discussion.}

\begin{table}[!thb]
\caption{\it Comparison of different strategies for selecting permutation $\phi$ and different number of training samples. DER and CI are shown on CH1-2spk.}\label{tab:train_mu_sample}
\vspace{-2mm}
\centering
\scalebox{.9}{
\begin{tabular}{rcccccc}
\toprule
& \multicolumn{6}{c}{Permutation based on}\\
\cmidrule{2-7} 
 & \multicolumn{3}{c}{sampling $z$}  & \multicolumn{3}{c}{$\mmu$}\\
\cmidrule(lr){1-1} \cmidrule{2-4} \cmidrule(lr){5-7}
\# of samples  & 1 & 12 & 100 & 1 & 12 & 100\\
\midrule
DER (\%) & 8.06 & 7.92 & 7.84 & 7.88 & 8.12 & 7.83\\
CI & \footnotesize{$\pm$1.03} &  \footnotesize$\pm$0.97 &  \footnotesize$\pm$1.01 &  \footnotesize$\pm$1.02 &  \footnotesize$\pm$1.01 &  \footnotesize$\pm$1.00\\
\bottomrule
\end{tabular}
}
\end{table}

\begin{figure*}[!thb]
    \centering
    \begin{subfigure}[b]{\textwidth}
        \centering
        \includegraphics[width=\textwidth]{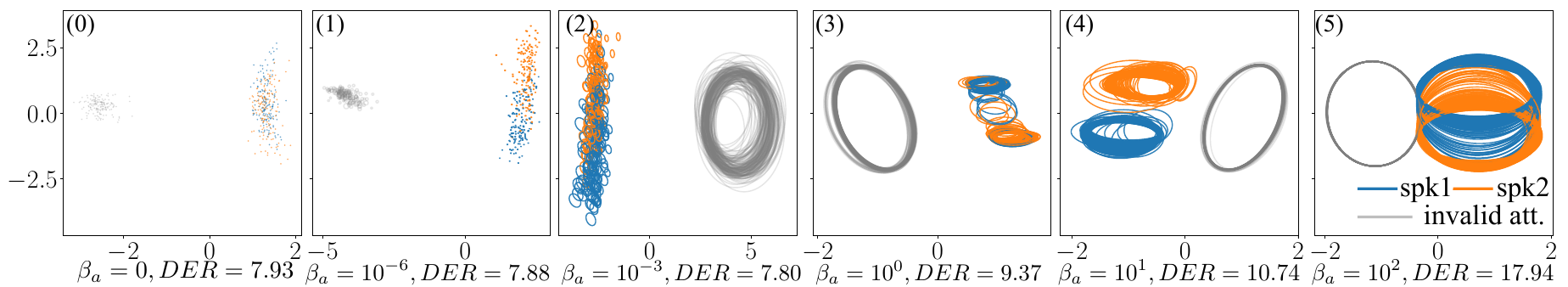}
       \vspace{-4mm}
        \caption{\it Visualizing attractor on $\beta_e=0$}
        \label{fig:visualization_gaussian_a}
    \end{subfigure}
    \hfill
    \begin{subfigure}[b]{\textwidth}
        \centering
        \includegraphics[width=\textwidth]{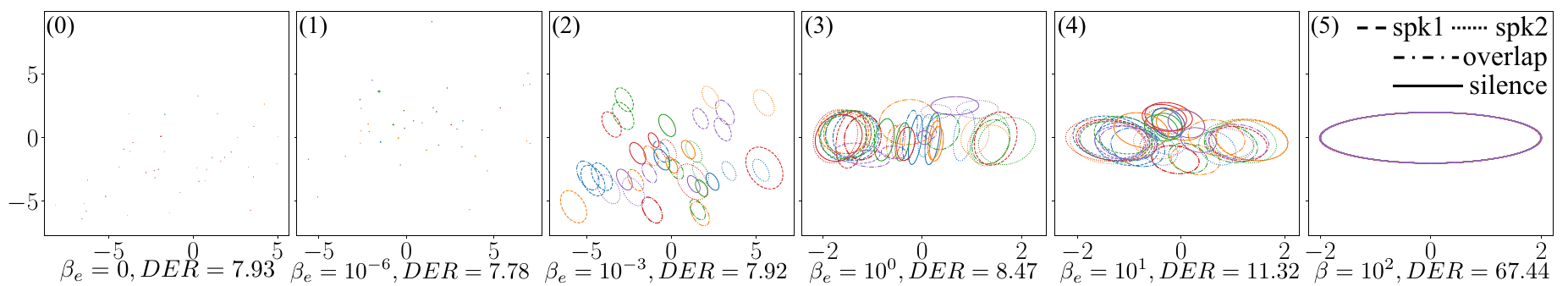}
        \vspace{-4mm}
        \caption{\it Visualizing frame embedding on $\beta_a=0$}
        \label{fig:visualization_gaussian_e}
    \end{subfigure}
    \caption{\it Visualization of attractors and frame embeddings as Gaussian distributions, after projecting them to two dimensions using PCA (DERs are in \%). In subfigure (b), colors represent different audio files; line styles denote individual speakers, overlap, and silence. Note that in subfigures (0) the attractors and frame embeddings are deterministic and are therefore represented by dots.}
   \label{fig:visualization_gaussian}
\end{figure*}

\subsection{Visualization of attractors and frame embeddings}
\label{sec:visualization}

\begin{figure}[h]
    \centering
    \includegraphics[width=\columnwidth]{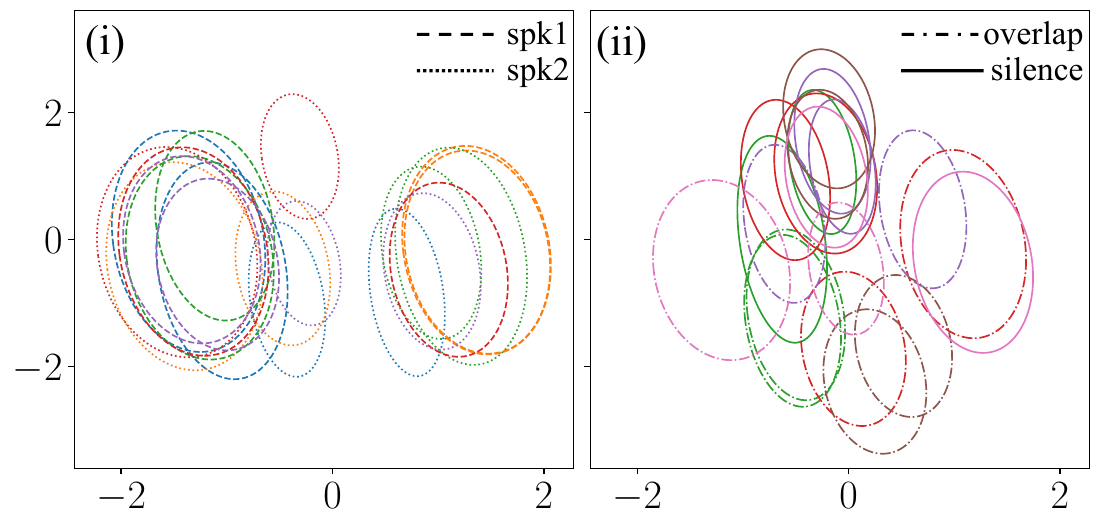}
    \vspace{-7mm}
    \caption{\it Decomposed visualization of Fig.~\ref{fig:visualization_gaussian_e}.4 ($\beta_a=0, \beta_e=10^{1}$). (i) two single speakers, and (ii) overlap and silence.}
     \label{fig:gaussian_singlespk_silolp}
\end{figure}

To get insights into what key information is encoded in the frame and speaker representations of the EEND-EDA model, we visualize the posterior distribution of the attractors $p(\zz_s|\xx) = \mathcal{N}(\mmu_s, \diag(\ssigma_s^2)$), and the frame embeddings $p(\zz_t|\xx) = \mathcal{N}(\mmu_t, \diag(\ssigma_t^2)$).
For the visualization, we project the attractor/embeddings into a two-dimensional space using principal component analysis (PCA) and then plot them as Gaussian ellipses, as shown in Fig.~\ref{fig:visualization_gaussian}.
Each ellipse shows the contour of the Gaussian distribution representing the corresponding stochastic enconding projected to the two dimensional space using PCA. The contour corresponds to the distance of two times the standard deviation from the mean.

\begin{figure}[!t]
    \centering
    \vspace{-1mm}
    \includegraphics[trim={0 0 0 0.1cm},clip,width=\columnwidth]{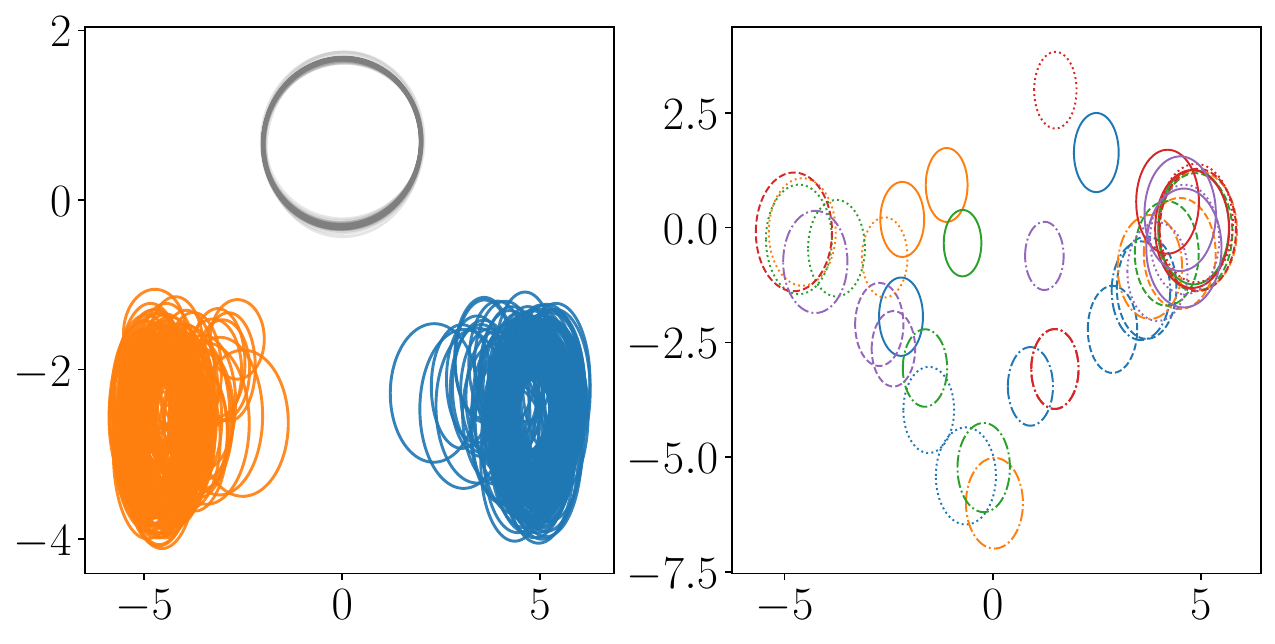}
    \vspace{-7mm}
    \caption{\it Visualization of attractors (left) and frame embeddings (right) when \it $\beta_a = \beta_e=10^{1}$.}
     \label{fig:gaussian_ze_1e1}
     \vspace{-3mm}
\end{figure}

To visualize the attractors, we first consider a fixed $\beta_e=0$ and a wide range of $\beta_a$ weights as shown in Fig.~\ref{fig:visualization_gaussian_a}, to get a good understanding of the effect of the regularization. We used the entire CH1-2spk to generate the plots. For each recording, three ellipses are shown in the figure, one with each category: the first speaker (spk1), the second speaker (spk2), and the invalid attractor.
To visualize the frame embeddings, we equivalently consider $\beta_a = 0$ and a wide range of $\beta_e$, and randomly selected a few frames from five random audio samples from CH1-2spk (not to clutter the images). 
From left to right, Fig.~\ref{fig:visualization_gaussian} visualizes the model's behavior with varying weights from lower to higher (corresponding to lighter to stronger regularization). 

As expected, Fig.~\ref{fig:visualization_gaussian} demonstrates the model's transition from deterministic to stochastic as the weights change. When the weight is zero (as in Fig.~\ref{fig:visualization_gaussian_a}.0 and Fig.~\ref{fig:visualization_gaussian_e}.0), the model is effectively the same as the original deterministic EEND-EDA.
When the weight is larger than $0$, yet close to zero, such as $10^{-6}$ in the Fig.~\ref{fig:visualization_gaussian_a}.1 and Fig.~\ref{fig:visualization_gaussian_e}.1, the attractor and frame embedding ellipses are close to point estimates, still similar to a deterministic model. 
With larger weights, more input information is forgotten, and we observe a shift towards a more regularized model characterized by the increased variance of the attractor and frame embedding distributions.

For the attractors, we can see that the valid speakers (blue and orange) are always well discerned from the invalid one (gray). 
When lower values of regularization are used, the attractor representations are spread across one of the axes (Fig.~\ref{fig:visualization_gaussian_a}.1-\ref{fig:visualization_gaussian_a}.2). 
Still, the representations cluster by the corresponding (first or second) decoded attractors and nothing suggest that they would be clustered according to the specific voice of corresponding speakers (which should be different for each input conversation). 
As $\beta_a$ increases, we observe that the distributions for all the first attractors (blue ones) significantly overlap, as seen in Fig \ref{fig:visualization_gaussian_a}.4. And the same thing happens for the second attractors (orange ones).
This has an impact on performance, degrading from 7.93\% DER in \ref{fig:visualization_gaussian_a}.0 to 10.74\% DER in \ref{fig:visualization_gaussian_a}.4. Still, the system performs reasonably well, while it is obvious that the attractors are only used for \textit{counting} the speakers, and no longer convey any speaker information.
In the last subfigure, Fig.~\ref{fig:visualization_gaussian_a}.5, we observe that the representation of both valid speakers is starting to collapse into a single mode, now significantly harming system performance.

\begin{figure*}[!t]
    \centering
    \includegraphics[width=1.98\columnwidth]{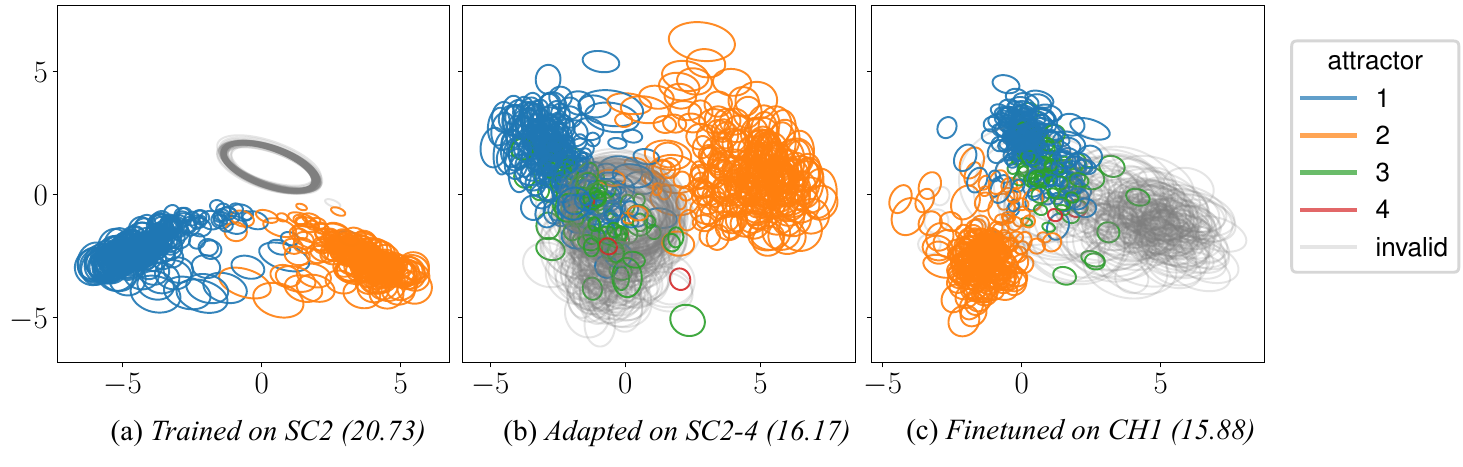}
    \vspace{-4mm}
    \caption{\it Visualization of Gaussian attractors on CH2 after training, adaptation, and fine-tuning. The colors denote attractors for valid or the first invalid (gray) attractor as predicted by the model. DERs (\%) are shown in brackets.}
    \label{fig:gaussian_CH2_train_adap_ft}
    \vspace{-3mm}
\end{figure*}

For the frame embeddings, we observe a similar trend for frame embeddings of a single speaker. 
For clearer visualization, we zoomed into Fig.~\ref{fig:visualization_gaussian_e}.4 and decomposed it into two subfigures presented in Fig.~\ref{fig:gaussian_singlespk_silolp}.
In Fig.~\ref{fig:gaussian_singlespk_silolp}.(i), shows the distributions of frame embeddings where only one of the speakers is active. We can observe that speakers within the same audio (in the same color) are generally distinguishable. For example, green dashed lines (corresponding to the first attractor) on the left are far from the green dotted lines (corresponding to the second attractor) on the right. 
However, distributions of speakers across audios are overlapped and generally clustered by the corresponding attractors. That indicates that frame embeddings only require enough information to discriminate between speakers of the same audio (intra-audio), but do not need to encode individual identities across different audio (inter-audio).
Besides, as shown in Fig.~\ref{fig:gaussian_singlespk_silolp}.(ii), silence across varying audios are clustered, and they are distinguishable from speech activity to some extent. This indicates that the frame-embedding requires encoding the information for speech activity. This aligns with the initial fundamental concept of the EEND model, which combines voice activity detection with diarization. Finally, we can also observe that, as expected, overlap frame representations are spread across the space between the single-speaker representations.

Equations \eqref{eq:KLD_loss_e} and \eqref{eq:KLD_loss_a} assume that the variational approximation to the marginal encoding distribution (see $r(\zz)$ in equation~\eqref{eq:vib_loss2}) is fixed to be standard normal. We use this configuration in all our experiments, although, it generates a problem when both frame embeddings and attractors ($\zz_t$ and $\zz_s$) are simultaneously regularized using the KLD loss. 
When only one type of encodings (either $\zz_t$ or $\zz_s$) is regularized, as in Fig.~\ref{fig:visualization_gaussian} the encoding distributions span range around $[-2, 2]$ for higher $\beta$ values. This range aligns with the standard normal Gaussian distribution that the KLD regularizes towards. However, as can be seen in Fig.~\ref{fig:gaussian_ze_1e1}, the encoding distributions span 3 times larger range, when regularizing both $\zz_t$ and $\zz_s$ simultaneously even with high regularization weights $\beta_a=\beta_e = 10^{1}$. 
Such a high dynamic range is enforced by the cross-entropy loss demanding high values of $p_{s,t}$, which can be achieved only with encodings $\zz_t$ and $\zz_s$ that are not close to the origin (see equation~\eqref{eq:eendeda_p}) and therefore violating the assumed standard normal marginal distribution.
This problem could be easily resolved by modeling the marginal distribution $r(\zz)$ by a Gaussian distribution with trainable parameters, which is left for future research.

\subsection{Analysis across training, adaptation, and fine-tuning}

In this section, we compare the performance of the proposed EEND-EDA with VIB in a real scenario with more speakers.
For these experiments, we considered the three optimal configurations of $\beta$ weights observed in Section \ref{sec:res_weight}, that is: (I) $\beta_e=0, \beta_a = 10^{-3}$, (II) $\beta_e = 10^{-6}, \beta_a = 0$, and (III) $\beta_e = \beta_a = 10^{0}$. Results on CH2 are shown in Table \ref{tab:res_veend_eda}.
We can see that VIB maintains diarization performance closely aligned with the baseline, even for the configuration of weights (III), for which the strong regularization filters speaker information, as was shown in Fig. \ref{fig:gaussian_ze_1e1}.
Still, we noticed that fine-tuning the EEND-EDA model with VIB on the real data results in a smaller improvement in the performance, as compared to the fine-tuning in the original EEND-EDA. 

To get some further insight into these results, in Fig.~\ref{fig:gaussian_CH2_train_adap_ft} we visualize the attractors of the EEND-EDA with VIB, when $\beta_a = \beta_e = 10^{0}$, after training stage, after adaptation to SC2-4, and after fine-tuning to CH1.
In the first step, the model is only trained with recordings of two speakers so it can only output two valid attractors. In this case, shown in Fig.~\ref{fig:gaussian_CH2_train_adap_ft}.(a), the 1\textsuperscript{st}, 2\textsuperscript{nd}, and the invalid attractors, are clearly separated. 
After adapting the model to expect a variable number of speakers, in Fig.~\ref{fig:gaussian_CH2_train_adap_ft}.(b), we can see that the 3\textsuperscript{rd} and 4\textsuperscript{th} valid attractors are sometimes decoded by the model.
However, these attractors partially overlap with the invalid attractor (gray), and the \emph{attractor clusters} become less compact.
Given the training (and evaluation) data, we can assume that the first two attractors will almost always be valid.
This is reflected in the plot: The 1\textsuperscript{st} attractor (blue) and the 2\textsuperscript{nd} (orange) are the most distinct ones. However, green and red ellipses overlap considerably with the gray (invalid) ones, revealing the higher uncertainty of the model about the 3\textsuperscript{rd} and 4\textsuperscript{th} attractors. This is intuitive since sometimes the 3\textsuperscript{rd} (or 4\textsuperscript{th}) attractor has to be considered invalid, thus, modeling them with closer representations to the invalid one might be beneficial.  
Observing plot (c), we can assume that fine-tuning could help to enhance the model’s discriminative capability against invalid speakers. The resulting plot shows that fine-tuning makes the invalid speaker more clearly separated from the newly discovered third and fourth valid speakers.
Nevertheless, it should be pointed out that in the plots we only use the first two dimensions after applying a PCA transformation. It is possible that, in other dimensions, the green and red ellipses do not overlap so much with the gray ones.

It is worth noting that the regions for the 1\textsuperscript{st}, 2\textsuperscript{nd}, 3\textsuperscript{rd} and 4\textsuperscript{th} attractors are fixed in the space, regardless of the input recording. This reinforces the conclusion from Section~\ref{sec:visualization}: the attractors serve as anchors for \textit{counting} speakers and do not seem to convey (much) speaker information.
Finally, the representations after the fine-tuning step, in Fig.~\ref{fig:gaussian_CH2_train_adap_ft}.(c), are only slightly different and even if some recordings contain more than four speakers, the small development set is not enough for the model to learn that more speakers should be detected.

\begin{table}[!thb]
\caption{\it Comparison of DER (\%) and CI on CH2.} \label{tab:res_veend_eda}
\centering\footnotesize  
\scalebox{1}{
\begin{tabular}{l@{\hskip 3mm}lll}
\toprule
\multicolumn{2}{c}{\textbf{Model}} & \textbf{w/o FT} & \textbf{w/ FT} \\
\midrule
\multicolumn{2}{c}{\textbf{Baselines}} & & \\
\multicolumn{2}{l}{EEND-EDA\cite{horiguchi_encoder-decoder_2022}}  &   -    & 15.29\\
\multicolumn{2}{l}{EEND-EDA (ours)}        & 16.25 ($\pm$2.30)  & 15.72 ($\pm$2.34) \\
\multicolumn{2}{l}{EEND-EDA + 4FC}    & 16.74 ($\pm$2.36) & 15.50 ($\pm$2.21) \\
\midrule
\multicolumn{2}{c}{\textbf{EEND-EDA with VIB}} & & \\
$\beta_a = 0$ &  $\beta_e =10^{-6} $  & 16.06 ($\pm$2.34) & 15.81 ($\pm$2.33) \\
$\beta_a = 10^{-3}$ & $\beta_e =0 $& 16.58 ($\pm$2.35) & 16.04 ($\pm$2.17)  \\
$\beta_a = 10^{0}$ & $\beta_e =10^{0}$ & 16.17 ($\pm$2.40) &15.88 ($\pm$2.50) \\
\bottomrule
\end{tabular}
}
\end{table}

\section{Conclusion}
 This study introduced VIB into EEND-EDA to understand what information needs to be stored in the frame embeddings and attractors, so that the model is effective in the diarization task. Our analyses show that attractors do not require substantial speaker characteristic information to obtain reasonable performance. Similar conclusions can be drawn from the frame embeddings: encoding unique individual identities across different audio samples (inter-audio) is not imperative; rather, the information to discriminate speakers within the same audio (intra-audio) is enough for diarization.

 Even though intuitively speaker characteristic information should not be necessary to perform diarization (models only need to ``tell speakers apart'' within the audio), this is, to the best of our knowledge, the first analysis in the context of end-to-end diarization models that provides evidence in this direction. We showed that models can perform reasonably well even when introducing strong regularization terms (based on VIB) for internal representations. Note, that this same strategy could be applied to produce more privacy-aware models.

\section{Acknowledgements}
The work was supported by the Czech Ministry of Interior project No. VJ01010108 ``ROZKAZ'', Horizon 2020 Marie Sklodowska-Curie grant ESPERANTO, No. 101007666, SOKENDAI Student Dispatch Program, and Japan Science and Technology Agency Grants JPMJFS2136. Computing on IT4I supercomputer was supported by the Czech Ministry of Education, Youth and Sports through the e-INFRA CZ (IDs 90140 and 90254).

\bibliographystyle{IEEE}
\balance
\bibliography{main}
\onecolumn

\begin{appendices}
\section{Appendix}

\subsection{Visualization for model trained on SC2 with varying weights $\beta_a = \beta_e \ge 0$}

\begin{figure*}[!hb]
    \centering
    \vspace{-1mm}
    \includegraphics[width=0.94\textwidth]{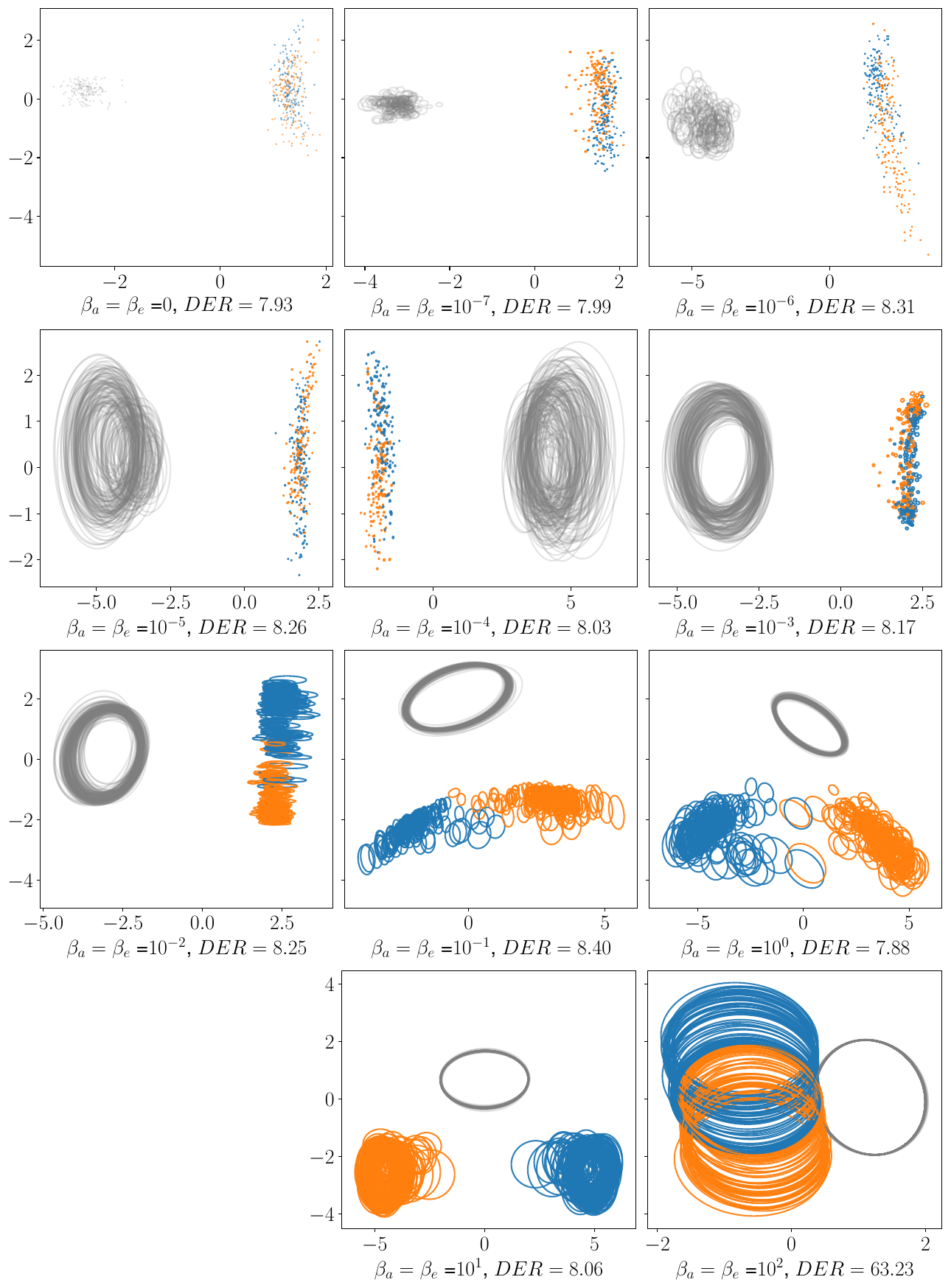}
       \vspace{-4mm}
    \caption{Visualization of Gaussian attractors on $\beta_a = \beta_e \ge 0$, after projecting them to two dimensions using PCA (DERs are in \%).}
\end{figure*}
\newpage
\begin{figure*}[!hb]
    \centering
    \vspace{-5mm}
    \includegraphics[width=0.94\textwidth]{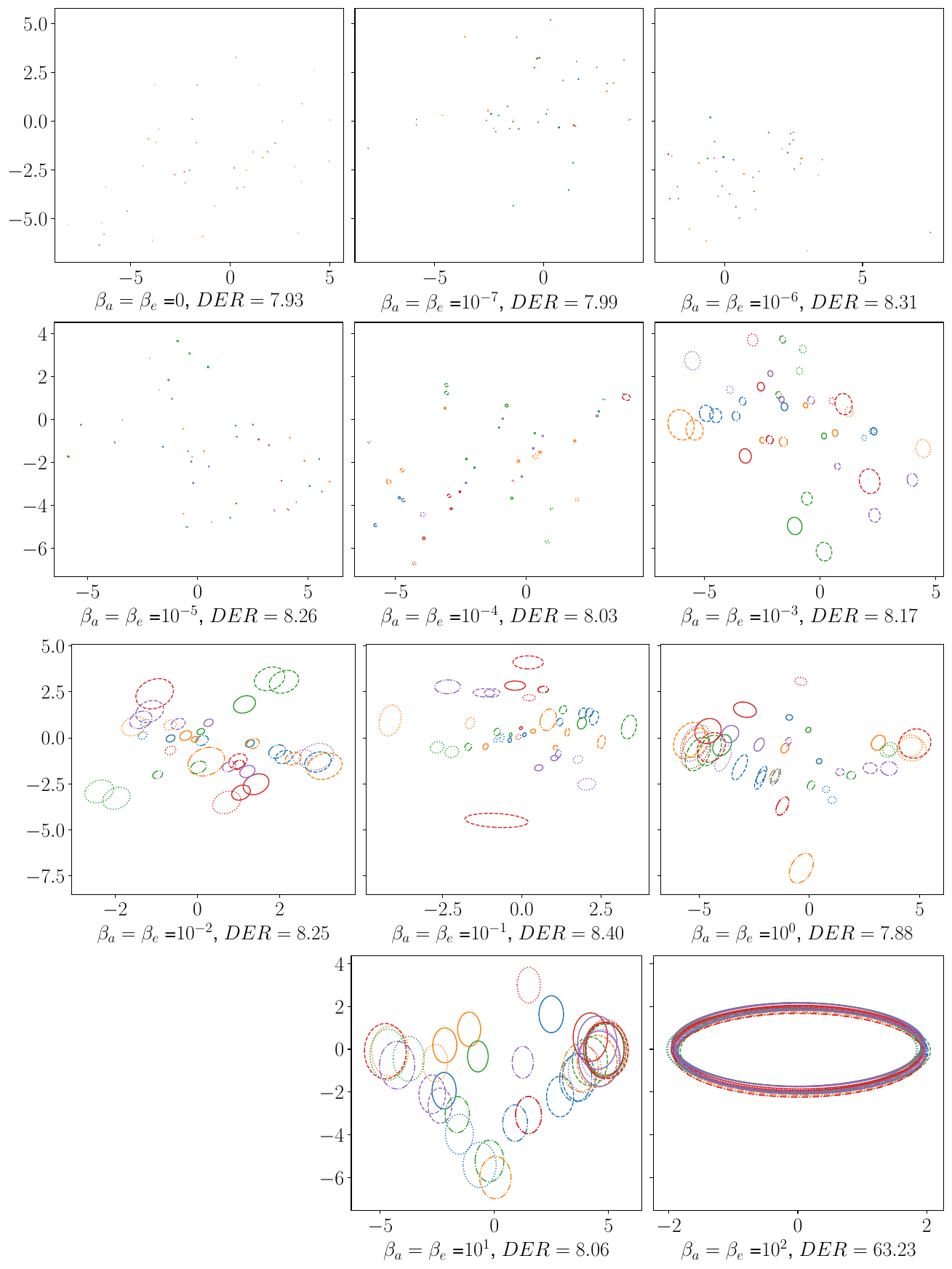}
       \vspace{-4mm}
    \caption{Visualization of Gaussian frame embeddings on $\beta_a = \beta_e \ge 0$, after projecting them to two dimensions using PCA (DERs are in \%).}
\end{figure*}

\newpage
\subsection{Visualization for model trained on SC2 with varying weights $\beta_a \ge 0, \beta_e = 0 $}

\begin{figure*}[!hb]
    \centering
    \vspace{-1mm}
    \includegraphics[width=0.94\textwidth]{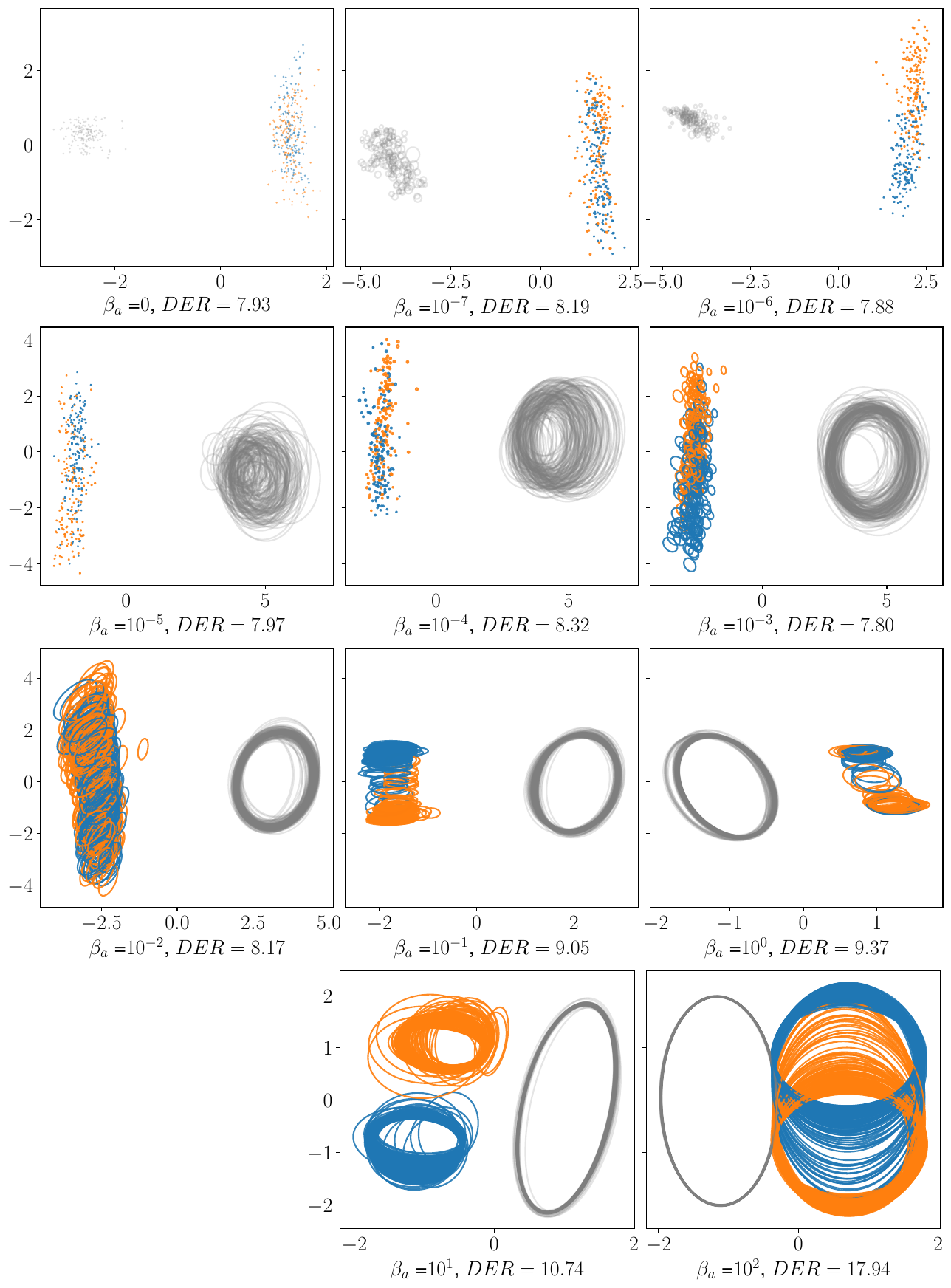}
       \vspace{-4mm}
    \caption{Visualization of Gaussian attractors on $\beta_a \ge 0, \beta_e = 0 $, after projecting them to two dimensions using PCA.}
\end{figure*}

\newpage
\subsection{Visualization for model trained on SC2 with varying weights $\beta_a = 0, \beta_e \ge 0 $}
\begin{figure*}[!hb]
    \centering
    \vspace{-1mm}
    \includegraphics[width=0.94\textwidth]{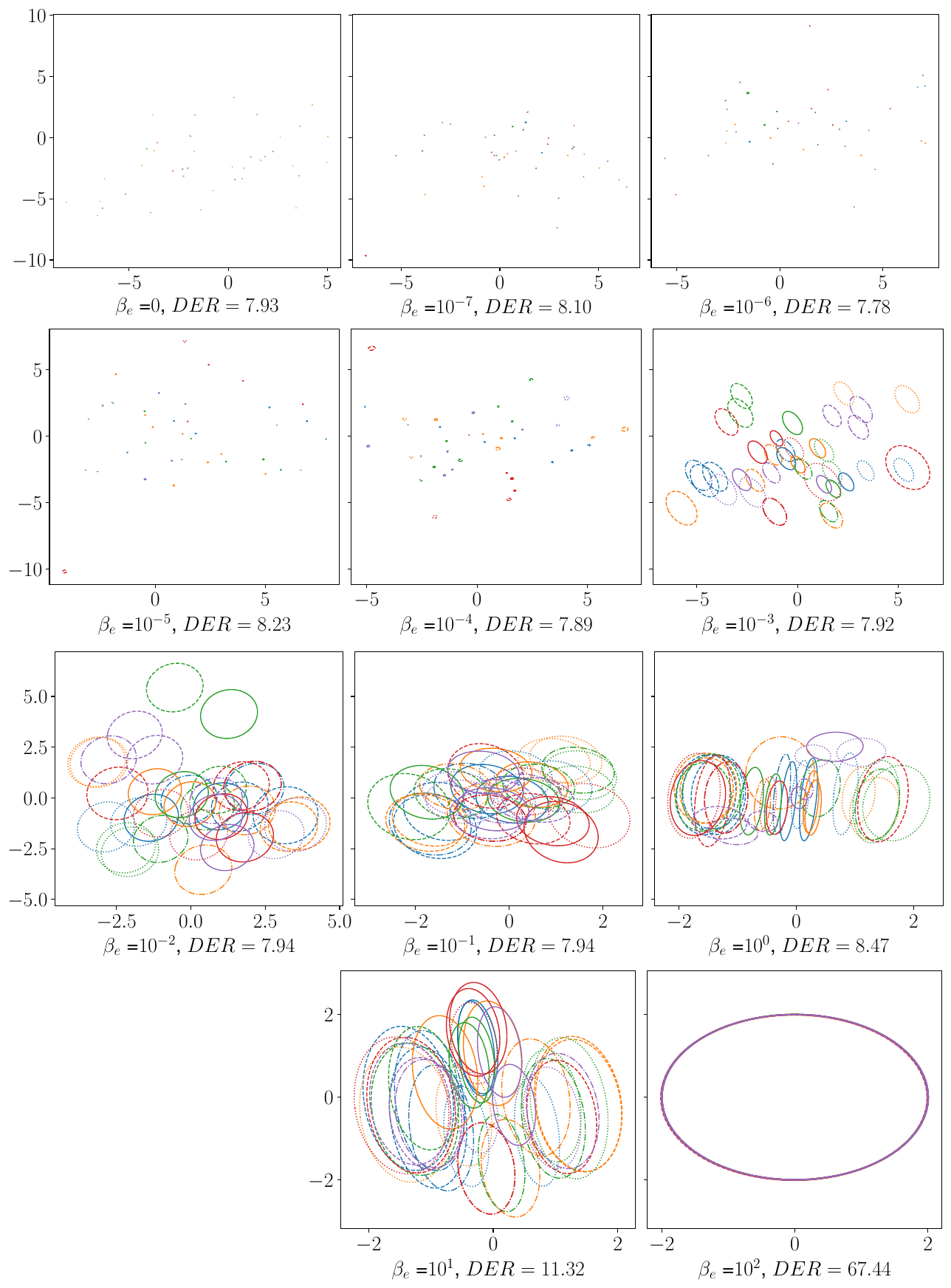}
       \vspace{-4mm}
    \caption{Visualization of Gaussian frame embeddings on $\beta_a = 0, \beta_e \ge 0 $, after projecting them to two dimensions using PCA.}
    
\end{figure*}

\end{appendices}

\end{document}